\documentclass[12pt]{article}
\usepackage{graphicx}

\title{Formation of an order in a system of exciton condensed phase islands in quantum wells}
\author{V. I. Sugakov \\  (Institute for Nuclear Research,
  Kiev, Ukraine)
  }  \date{}
\begin{document}
\maketitle

A theory of exciton condensed phase creation in two-dimensional system is presented.
    The theory is applied to explain the appearance of  the periodical fragmentation which
    was observed last years in luminescence from the ring around laser spot in crystal
     with double quantum wells.

     PACK: 71.35. Lk

\bigskip
    An electrons and holes of indirect excitons in semiconductor coupled quantum wells are localized
        in different wells and  as the result the excitons have long lifetime. Such system
        present interest for
         study of exciton condensation phenomena \cite{Tim}. Recently interesting experiments
        were fulfilled for indirect excitons in AlGaAs and InGaAs crystals with coupled quantum
    wells.   In  works
         \cite{But1,Snok1,But2,Snok2,But3,Snok3} the authors have
          observed a ring structure in emission of indirect excitons outside of laser spot.
          The distance between ring  and center of the excitation spot growths with
          increasing the pumping and  can reach several hundred microns that is much
          larger than the exciton diffusion length.
           At low temperature ($T\leq 2K$) the external ring is fragmented into periodical
           structure over
            macroscopic lengths.
    The mechanism of the luminescence ring formation  was proposed in \cite{But3,Snok3}.
     The origin of fragmentation along the ring is unclear. In present work the appearance
     of this fragmentation is explained by creation of
islands
      of exciton condensed phases along the ring. In \cite{Sug1,Sug2} it was shown that
      in the presence of an
       attractive interaction between excitons a high density exciton system  is unstable
        with respect to  the periodical structure appearance.
To obtain the spacial dependence of the exciton generation rate we shall use the
suggestion of paper
 \cite{But3,Snok3} that in absence of the light irradiation the quantum well contains electrons
  and  that holes created by light  are captured by
 quantum well more effectively than electrons. As the result the charge separation takes place
  from positive charge value  in the center of laser spot till
    negative one  far from the spot. The emission of light occurs from a ring
     where product of the electron and hole densities has maximum.
For determination of the density distribution we have used the system of equations
similar to that investigated in \cite{But3,Snok3} but we added the equation for the
exciton density. Lets $n_e, n_h, n_{ex}$ designate  the well densities of electrons,
holes and excitons, respectively. These values  satisfy to the following kinetic
equations
\begin{equation}
\frac{\partial n_e}{\partial t}=D_e\triangle_2
n_e+K_e(r)-Wn_en_h-\frac{n_e-n_0}{\tau_e},\label{1} \end{equation}
\begin{equation}
\frac{\partial n_h}{\partial t}=D_h\triangle_2
n_h+K_h(r)-Wn_en_h-\frac{n_h}{\tau_h},\label{2} \end{equation}
\begin{equation}
\frac{\partial n_{ex}}{\partial t}=D_{ex}\triangle_2
n_e+G-\frac{n_{ex}}{\tau_{ex}},\label{3} \end{equation}
                      where  $D_e$, $D_h$ and $D_{ex}$ are the diffusion
coefficients,  $\tau_e$, $\tau_h$ and $\tau_{ex}$ are the lifetime for electrons, holes
and excitons, respectively, $W$ is the  electron-hole recombination rate, $K_e(K_h)$ is
the electron (hole) creation rate in the well, $G$ is the exciton production rate
($G=qWn_en_h, q\leq 1$).

     Fig.1 shows the result of calculations of electron, hole and exciton  densities for
    the  following values of parameters: $D_e=200cm^2/s$,
      $D_h=50cm^2/s$,$D_{ex}=10cm^2/s$, $W=150cm^2/s$ , $q=0.9$,
     $\tau_e=\tau_h=10^{-5}s$, $\tau_{ex}=10^{-7}s$, the spacial
distribution of the pumping  was approximated by the Gaussian curve with
     width $60\mu m$ and $K_h=2K_e$.  The qualitative picture of
        behavior of $n_e, n_h$   as a function of $r$ is similar to
results obtained in \cite{But3,Snok3}. The  position of the exciton density maximum
 determines the ring luminescence in experiments. We shall use the
 obtained value of exciton density to study the fragmentation of
 the exciton condensed phase. It is suggested that the condensation
 occurs due to the exciton-exciton interaction  and it is not the
  Bose-Einstain condensation in  the  wave  vector space with
  $\textbf{k=0}$.  We have  estimated the Van der Waals attraction
  using the well known formula of quantum mechanics and showed
  that  Van der Waals interaction between two indirect excitons in
     coupled  quantum wells exceeds the dipole-dipole interactions in the
     range of distances between excitons of order of several
      exciton radiuses (See Appendix). The   attractive interaction may cause
      the existence of the exciton condensed phase. Some models of
      phase transitions in a system of indirect excitons were
       studied in \cite{Loz}. Afterwards we shall suggest that
        condensed phase exists and it is described by some parameters which will be determined
        later.
\begin{figure}%1
\centerline{\includegraphics[width=4.0in]{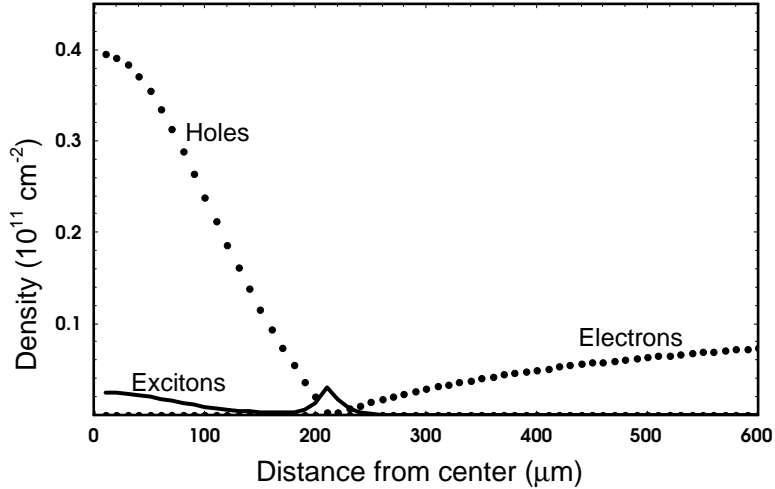}} \caption{The dependence of
electron, hole  and exciton  densities on distance from center} \label{fig:01}
\end{figure}
  Due to a finite value of the exciton lifetime the sizes of exciton
condensed phases are restricted.  As the  result in the two-dimensional case the
condensed phase  must exist as a system of  islands similar to electron-hole drops in
bulk semiconductors. In studied system the islands should be localized on the ring
where the exciton density has maximum.

\begin{figure}%2
\centerline{\includegraphics[width=4.0in]{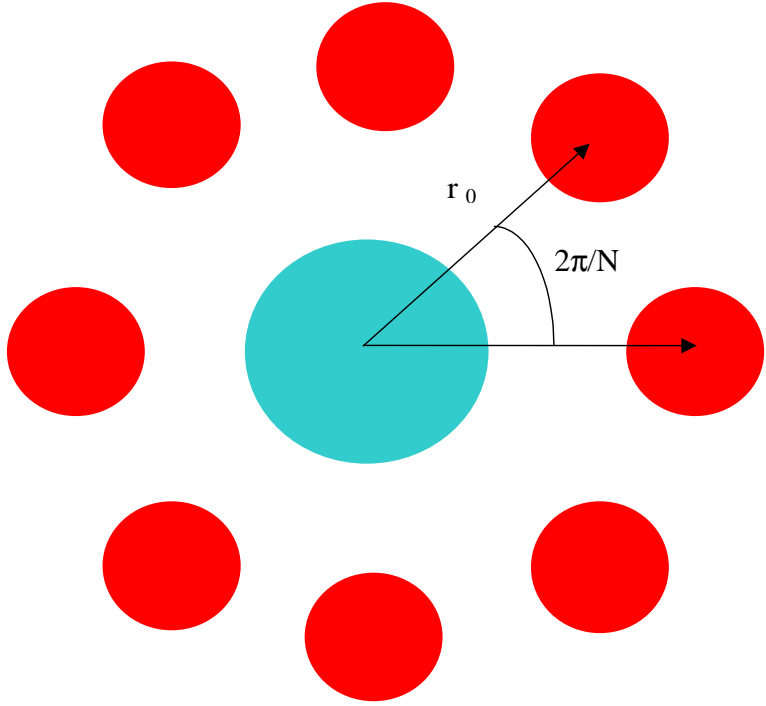}} \caption{The position of exciton
condensed phase islands around the laser excitation spot} \label{fig:02}
\end{figure}

Under following consideration we shall apply the theory of creation of the exciton
condensed phase islands in two-dimensional case \cite{Sug3} to studied system with
nonuniform pumping. Firstly, we shall study the system of condensed phase islands
periodically situated along the ring at maximum of the exciton density  with $r=r_0$
(Fig.2).
 $\varphi_m=2\pi m/N$ is the angle of $m$-th island on the ring. Afterwards we shall show
 that deviations (fluctuations) from
  periodicity are small. Let us consider the formation of some island, for example,the
  island with $m=0$. We introduce the distribution function $f_n$, which determines
  a probability of the island with $m=0$ to have $n$  excitons. The size of disks is
  determined by
  four processes: creation of excitons by pumping, capture of the excitons from environment,
  escape of excitons from the disk, and the exciton decay. The kinetic equation for the
  distribution
  function have the following form
  \begin{equation}
  \frac{\partial f_n}{\partial t}=-j_{n+1}+j_{n}, \label{4}
  \end{equation}
 where $j_n$  is the probability current
  \begin{equation}
  j_n=2\pi R_{n-1}W_{fi}(R_{n-1})c(R_{n-1})f_{n-1}-2\pi R_nW_{if}(R_n)c_if_n-
  \pi R_n^2f_n/\tau_{ex}+\pi R_{n-1}^2 \bar G_{n-1}f_{n-1}, \label{5}
  \end{equation}
  $R_n$  is the radius of the disk with $n$  excitons, $W_{fi}(R_n)$ and $W_{if}(R_n)$ are
  probabilities for the  exciton to be captured by the disk and to escape from the disk per
  unit length of
  the circle, respectively, $\bar G_{n}=\int G(r)dS_n/\pi R_n^2$ is the mean value of the
  exciton pumping
  over island area, $c(R_n)$ and $c_i$ are
  the concentrations of excitons on the circle of the disk and inside disk ($c_i=1/s_o$,$s_o$
   is the area
  per single
  electron-hole pair in the island). The exciton concentration $c(r)$  differs from $n_{ex}(r)$
  presented in Fig.1
  due to presence of islands which perturb the exciton concentration field around them.
    The following condition between transition probabilities  $W_{fi}$ and $W_{if}$ takes
    place due to the detail balance principle
       \begin{equation}
  \frac{W_{if}(R)}{W_{fi}(R)}=\frac{W{if}(\infty)}{W_{fi}(\infty)}\exp\left(\frac{a_2}{R}\right), \label{6}
  \end{equation}
where $W_{fi}(\infty)$  and $W{if}(\infty)$  are the transition probabilities in the
case of straight line boundary between condensed and gas phases,
$W_{fi}(\infty)/W{if}(\infty)=c_i/c_{\infty}$, $c_{\infty}$ is the equilibrium
concentration of excitons for the straight line boundary between condensed and gas
phases
     \begin{equation}
  c_{\infty}=c_{10}\exp (-\varphi/\kappa T), \label{7}
  \end{equation}
  $\varphi$ is the condensation energy per an exciton, $c_{10}=\gamma (m^*\kappa T/2\pi \hbar^2$,
  $m^*$ is the effective exciton mass,$\gamma$ is the degeneracy of the exciton state,
  $a_2=a_ls_o/\kappa T$, $a_l$ is the energy per unit of the disk length.

The connection between islands occurs due to dependence of the exciton  surface
concentration of considered island $c(R_n)$ in the kinetic equation (\ref{4}) versus
presence of other islands. We  suggest that the distance between disks is larger than
the disk radius  and the concentration field created by some island slowly changes in
limits of a size of the considered island.  As the result the exciton concentration on
the disk with $m=0$ may by presented in the form
 \begin{equation}
  c_{ex}(R)=\bar n_{ex}(r_0)+a_0 K_0((R/l)+\sum_{\mu\neq 0}K_0(2r_0\sin (\varphi_\mu/2)/l),
   \label{8}
  \end{equation}
where $K_0(x)$ is the modified Bessel function, $l$  is the diffusion length of free
exciton.
 Coefficients $a_\mu$  are determined by boundary conditions on every disk: the current of
excitons to an island should be equal difference between number of excitons captured by
the island and the number of excitons which escape from the island
    \begin{equation}
  2\pi RD_{ex}\frac{\partial c_{ex}(R)}{\partial R}=2\pi R(W_{fi}c_{ex}(R)-W_{if}c_i),
   \label{9}
  \end{equation}
Due to the system symmetry  all islands have
     equal parameters: $a_\mu=a_0$  for every $\mu$.
    Further we introduce the radius distribution function$f(R)=f_ndn/dR=2\pi Rf_n/s_o$.
    In steady case at $n>>1$ the solution of the
    equation (\ref{4}) has the form
          \begin{equation}
  f(\tilde R,N)=f_0\exp(-F(\tilde R,N),
   \label{10}
  \end{equation}
where
  \begin{equation}
 F(\tilde R,N)=-4\pi\int\limits_0^{\tilde{R}} \frac{2\nu \left(\tilde n_{ex}(r_0)-
 \tilde c_{10}e^{-\frac{(\varphi-a/\tilde R)}{T}}\right)-
 \tilde R\left(1-\tilde n_G+
 \frac{\nu (K_0(\tilde R/\tilde l)+\sigma (N))}{\tilde lK_1(\tilde R/\tilde l)}\right)}
 {2\nu \left(\tilde n_{ex}(r_0)+
 c_{10}e^{-\frac{(\varphi-a/\tilde R)}{T}}\right)-
 \tilde R\left(1+ n_G+\frac{\nu (K_0(\tilde R/\tilde l)+\sigma (N))}
 {\tilde lK_1(\tilde R/\tilde l)}\right)}\tilde Rd\tilde R,
   \label{11}
  \end{equation}
    \begin{equation}
  \sigma (N)=\sum_{\mu=1}^{N-1} K_0(2\tilde r_0\sin (\varphi_\mu/2)/\tilde l),
   \label{12}
  \end{equation}
Here we have introduced the dimensionless variables
     $\tilde R=R/\sqrt{s_o}, c(\tilde R)=c(R)/s_o, \nu=W_{fi}\tau_{ex}\sqrt{s_o},
     \bar n_G=\bar Gs_o\tau_{ex}$,
  $\varphi$ and $a$   are expressed in (\ref{10}) and (\ref{11}) in units of temperature.
The most probable radius is determined from the condition
    \begin{equation}
  \frac{\partial F(\tilde R,N)}{\partial\tilde R}=0,
   \label{13}
  \end{equation}

The mean radius depends on the number of islands is $\bar R=R(N)$. The probability for
 system to have  $N$ islands with radiuses $\tilde R_1, \tilde R_2,...\tilde R_N$
equals to
\begin{equation}
 W(N,\tilde R_1, \tilde R_2,...\tilde R_N)\approx \exp (-\sum _{i} F(\tilde R_i,N).
   \label{14}\end{equation}
After integrating over the radiuses of the disks we obtain the probability for the
system to have $N$ islands on the ring
            \begin{equation}
 W(N)= \exp (-\Phi (N)),
   \label{15} \end{equation}
where
    \begin{equation}
\Phi (N)=-N\ln z(N), z(N)=\int_0^\infty \exp (-F(\tilde R(N),N) d\tilde R.
   \label{16}\end{equation}

The number of islands is determined by the condition
      \begin{equation}
  \frac{\partial\Phi}{\partial N}=0.
   \label{17} \end{equation}

The formulae  (\ref{13}, \ref{17})  determine the mean radius and number of islands.
Since islands of condensed phases  derive excitons from the same source  two islands
can not be situated close one to another. Moreover, the distance between islands can
not be large because in this case the exciton density between them becomes greater than
critical value and as the  result the new island may appear. So, there is specific
interaction between condensed phases trough the exciton concentration fields. As the
result the dependence $\Phi (N)$ (\ref{17}) has minimum at some value $N=N_s$ that
determines the number of islands on the ring. We have calculated the some parameters of
the islands on the ring using  the equations (\ref{10}-\ref{17}). The exciton
parameters and parameters of pumping are the same as parameters in Fig.1. The following
parameters of the condensed phase were chosen : $m^*=0.37m_0, \varphi=10K,a_2=18K,
s_o=10^{-11}cm^{2}$. It is seen from Fig.3 that the distance between islands ($d=2\pi
r_0/N_s$) increases with rise of temperature, the island radius changes slowly with
temperature. At considered parameters the condensed phase exists till $2.16K$.

\begin{figure}%3
\centerline{\includegraphics[width=4.0in]{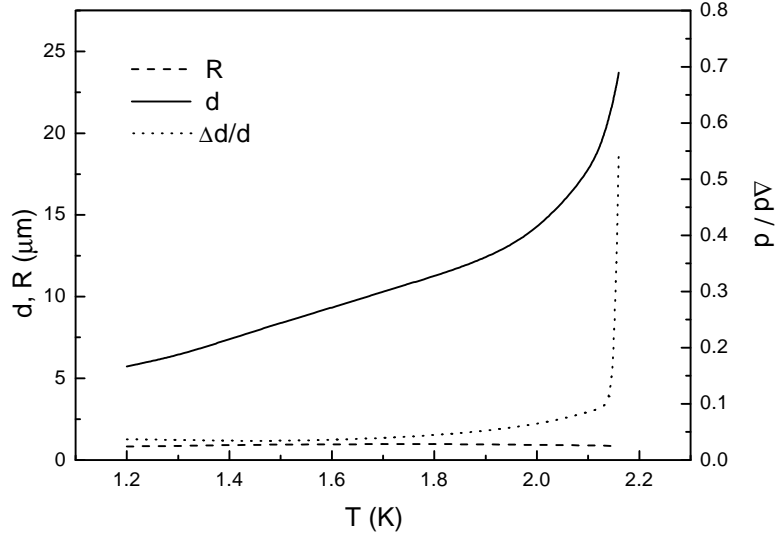}} \caption{The dependence of  the
island radius (R),   the distance between islands (d) and the ratio of island
fluctuation shift to distance between islands  ($\Delta d/d)$ on temperature}
\label{fig:03}
\end{figure}
  To
estimate the island position fluctuation we studied  the shift of some island versus
the angle $\varphi$ along the ring at fixed position of  all other islands (similar to
Einstein model  for crystal oscillations). In this case the function $\sigma (N)$  in
(\ref{13}) depends on $\varphi$  ($\sigma (N)\rightarrow \sigma (N,\varphi$)) and the
function $f(\tilde R,N)\rightarrow f(\tilde R,N,\varphi)$ determines the angle
distribution function. Using its we calculated the mean fluctuation shift of island
along the ring, which equals $\Delta d=r_0\sqrt{\overline{\varphi^2}}$.  It is seen
from Fig.3 that the relative shift $\Delta d/d$ is small ($\Delta d/d<<1$ ) and
increases only in the vicinity of temperature at which the condensed phase disappears.
By similar way using (15) the relative fluctuations of the island number ($\Delta
N/N_s=\sqrt{\overline{(N-N_s)^2}}/N_s)$ were calculated. This value is much less than
unity. For example, we have $\Delta N/N_s=7\cdot 10^{-3}$  at $T=2K$. Thus, the studied
system has perfect periodicity. These results explain the periodical fragmentation
observed in \cite{But1}.

 We suggested that outside of  islands the
exciton gas is nongenerated. At considered parameters the exciton density at the
surface of disks outside of disks equals  to $7.2\cdot 10^{8} cm^{-2}$ and is
significantly lower than the critical Bose-Einstain condensation density. Inside of
islands the exciton density is significantly higher and taking into account the Bose
statistic
 may be important. But presented theory did not consider the model of the condensed
phase and needs only some parameters of the condensed phase:  energy and occupied
volume per electron-hole pair and surface energy. Moreover, the condensed phase may
present electron-hole liquid phase.

 In the paper the binding
energy per electron-hole pair in  condensed phase was chosen ($ \varphi =10K=0.8 meV$).
This value gives the critical temperature of fragmentation disappearance ($2.16K$)
close to experimental meaning ($\sim 2K$).The value  of $\varphi$ is less than width of
the fragment emission line ($1.3 meV$ \cite{But1}) and the  emission line spectral
positions of free exciton and the fragment coincide practically. Even there is very
small (much less than width) shift of the fragment emission to blue side (See Fig2c of
\cite{But1}). But mechanisms, that determine the width  (scattering by fluctuation
potential, phonons, between excitons and others), give different contribution to the
shapes of emission spectra (including to the shift of lines) for the free excitons and
the exciton condensed phase.  So, if the exciton binding energy to condensed phase is
less than the bandwidth,  the physical information from comparison of emission spectra
shapes of free excitons and the condensed phase can be obtained  only together with
analysis of line broadening mechanisms for these states. For that the detailed theory
of the condensed phase  is needed.

Acknowledgment. This research work was supported partially by INTAS grant No.03-51-5266
and by the Ukrainian Ministry of Education and Science (project No. 02.07/147).
\section*{Appendix}
To  estimate the Van der Waals interaction between two excitons let us  use the well
known formula
\begin{equation}
U_{vdw}=-\sum_{i_{1}\neq 0,i_{2}\neq 0} \frac{\mid <0,0 \mid U_{dd}\mid i_1,i_2> \mid
^2}{E_{i_1,i_2}-E_{0,0}}, \label{A1}
\end{equation}
where $U_{dd}=\frac{\bf{P}_1\cdot \bf{P}_2}{\varepsilon
R_{12}^3}-\frac{3\bf{P}_1\cdot\bf{R_{12}}\bf{P}_2\cdot\bf{R_{12}}}{\varepsilon
R_{12}^5}$ is the operator of the dipole-dipole interaction between excitons, $R_{12}$
is the distance between excitons, $|0,0>$ and $|i_1,i_2>$ are the ground and excited
states of a system with two excitons. The formula $U_{dd}$ can be used if the distance
between excitons is larger than exciton radius. For calculation we have used the
following model. Electron and hole are localized in different wells. The distance
between wells is equal to $\delta$. For estimation we took into account only the ground
state and two lowest excited states, the dipole transition matrix element of which from
ground state to excited one does not equal to zero.  The wave functions of such type
have the form
\begin{equation}
|0>=A_0 \exp (-\alpha _0 r); |1a>=A_{1a}r\cos \varphi\exp (-\alpha _{1a}r);
|1b>=A_{1b}r\sin \varphi\exp (-\alpha _{1b}r); \label{A2}
\end{equation}
$r$ is the distance between electron and hole in exciton in the plane of wells, $\alpha
_0 ,\alpha _{1a},\alpha _{1b}$ are variational parameters.
\begin{figure}%4
\centerline{\includegraphics[width=4.0in]{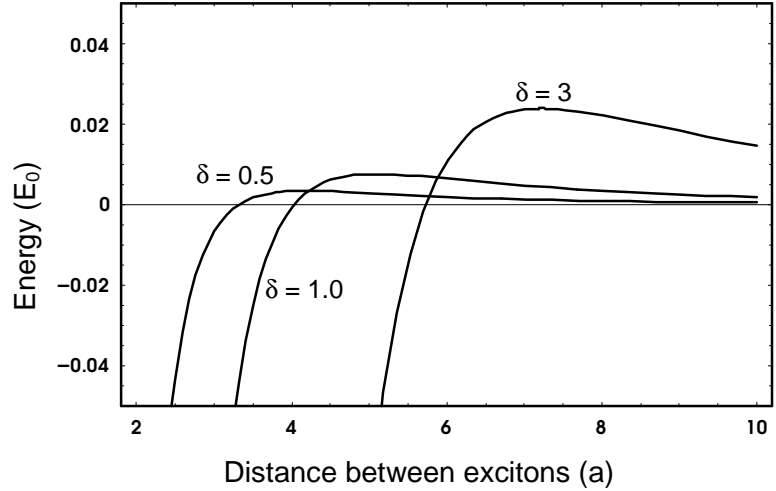}} \caption{The dependence of the sum
of the Van der Waals  and dipole-dipole interaction energies between two excitons in
coupled quantum wells on the distance between excitons for the different values of the
distance between wells $\delta$. $E_0$ and $a$ are the energy   and radius of exciton,
respectively, in bulk crystal} \label{fig:04}
\end{figure}
The results of calculation of the dipole-dipole and Van der Waals interaction
($<0,0|U_{dd}|0,0> +U_{vdw}$) are presented in Fig.4
 for different values of $\delta$. The energy of bulk exciton  is chosen as an unit of energy, radius of bulk exciton (the
value of order of $50A^o$) is the unit of length.  At the large distance the
interaction between excitons is dipole-dipole repulsive. But the attractive Van der
Waals interaction exceeds the dipole-dipole repulsion at distances less than $3\div 6$
exciton radiuses and total interaction is attractive.
 The taking into account the
other more highest excited intermediate states in (\ref{A1}) should lead to increasing
the presented estimations of Van der Waals interaction.  So,  Van der Waals interaction
is really larger. At small distances between excitons the exchange interaction becomes
important, also the approximation of dipole-dipole interaction becomes inapplicable. As
the result the total interaction is repulsive one at small distances.

It should be noted that attractive interaction may be significant at low temperature
when condensed phase is created.  At higher temperature  excitons are distributed
uniformly over crystal, the mean distance between excitons exceeds the region of
attraction  and main contribution to the exciton-exciton interaction gives
 the dipole-dipole one.


\begin{thebibliography}{99}
\bibitem{Tim}  A.V. Larionov ,  V.B. Timofeev. Pis'ma v Zh. Eksp. Teor. \textbf{73}, 342,
(2001).
\bibitem{But1} L.V. Butov , A.C. Gessard, D.S. Chemla. Nature \textbf{418}, 751, {2002}.
\bibitem{Snok1}  D. Snoke , S. Denev ,
Y. Liu , L. Pfeiffer , K. West.  Nature \textbf{418},  754, (2002).
\bibitem{But2}L.V. Butov. Sol.St.Comm. \textbf{127},
 89, (2003).
\bibitem{Snok2} D. Snoke ,Y. Liu, S. Denev ,L. Pfeiffer,K. West.   Sol.St.Comm. \textbf{127},
187,(2003).
\bibitem{But3}  L.V. Butov, L.S. Levitov,  A.V. Mintsev, B.D. Simons, A.C. Gossard ,
D.S. Chemla.  Phys Rev. Lett. \textbf{92},  117404, (2004).
\bibitem{Snok3}   R. Rapaport,  Yang Chen, D. Snoke , Steven H. Simon,
 Loren Pfeiffer,  Ken West, Y. Liu and  S. Denev.   Phys Rev. Lett. \textbf{92},  117405, (2004).
\bibitem{Sug1}   V.I. Sugakov.   Fizika Tverdogo Tela \textbf{21},
 562,(1986).
 \bibitem{Sug2}   V.I.Sugakov.   Phase Transition \textbf{75},  953, (2002).
 \bibitem{Loz} Yu.E. Lozovik , O.L. Berman.   Pis'ma v Zh. Eksp. Teor. Fiz \textbf{64}, 526,(1996).
 \bibitem{Sug3}  V.I. Sugakov.  Fizika Tverdogo Tela \textbf{46},1455, (2004); Physics
 of the Solid State \textbf{46},1496, (2004).

 \end{thebibliography}
\end{document}